\documentstyle[proceedings,epsfig]{crckapb}

\def\gs{\mathrel{\raise0.35ex\hbox{$\scriptstyle >$}\kern-0.6em 
\lower0.40ex\hbox{{$\scriptstyle \sim$}}}}
\def\ls{\mathrel{\raise0.35ex\hbox{$\scriptstyle <$}\kern-0.6em 
\lower0.40ex\hbox{{$\scriptstyle \sim$}}}}

\begin{opening}
\title{THE DIFFERENTIAL MAGNIFICATION OF HIGH-\protect\\
REDSHIFT ULTRALUMINOUS INFRARED GALAXIES} 

\author{A.\,W.\ Blain}
\institute{Cavendish Laboratory, Madingley Road, Cambridge, UK}

\end{opening}

\runningtitle{Differential magnification of ULIRGs}

\begin{document}

\section{Introduction}

The spectral energy distribution (SED) of an ultraluminous infrared galaxy 
(ULIRG) at wavelengths between about 10 and 1000\,$\mu$m is dominated by 
the thermal emission from interstellar dust, which is heated by absorbing blue 
and ultraviolet light from young stars and/or an active galactic nucleus (AGN). 
Mid- and far-IR radiation from galaxies at redshifts $z \ls 1$ has been detected 
directly by the {\it IRAS} and {\it ISO} space-borne telescopes. The redshifted 
dust emission from more distant galaxies can be detected efficiently at 
longer submillimetre wavelengths. The only one with a known redshift 
and a constrained mid-IR SED is SMM\,J02399$-$0136 at $z=2.8$ 
(Ivison et al.\ 1998).  
High-$z$ galaxies were also detected 
by {\it IRAS}, but only those with flux densities enhanced by gravitational 
lenses. Three such galaxies are known -- IRAS\,F10214+4724 (Rowan-Robinson 
et al.\ 1991), H1413+117 (Barvainis et al.\ 1995) and APM\,08279+5255 (Irwin et 
al.\ 1998). The SEDs of the four high-$z$ galaxies, and two well studied 
low-$z$ ULIRGs -- Arp\,220 and Markarian 231 (Klaas et al.\ 1997; Rigopoulou 
et al.\ 1996; Soifer et al.\ 1999) -- are shown in Fig.\,1. 

\begin{figure}
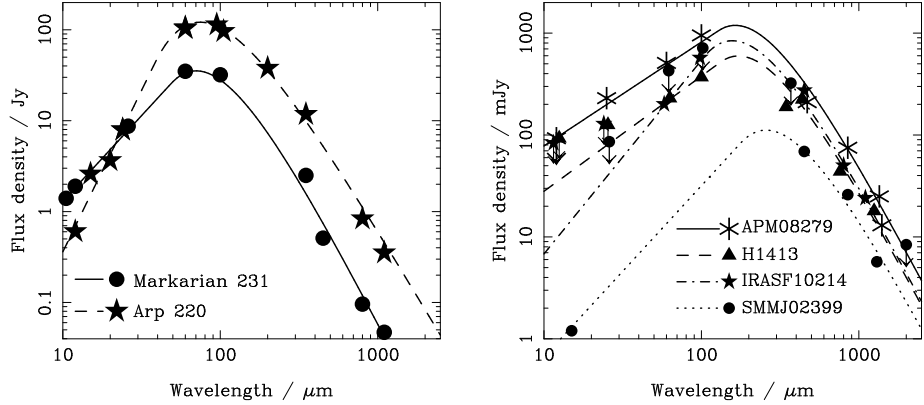

\vskip -3mm
\begin{center}
\centerline{\epsfig{file=blain_fig1a.ps, width=5.3cm, angle = -90} \hskip 5mm
\epsfig{file=blain_fig1b.ps, width=5.3cm, angle = -90}}
\hskip 5mm
\end{center} 
\vskip -3mm
\caption{The SEDs of two low-$z$ (left panel) 
and four high-$z$ (right panel) ULIRGs. Note that the mid-IR SEDs of 
APM\,08279+5255 (Lewis et al.\ 1998; Downes et al. submitted), H1413+117 and 
IRAS\,F10214+4724 are probably modified by differential magnification. 
The SEDs are described by a power law, $f_\nu \propto \nu^p$, at short 
wavelengths and by a blackbody spectrum at temperature $T_{\rm d}$, modified 
by a dust emissivity $\epsilon_\nu \propto \nu^\beta$ with $\beta \simeq 1$ at 
long wavelengths. The best fitting values of $T_{\rm d}$ and $p$ are listed in 
Table\,1.}
\end{figure}

\section{Spectral energy distributions} 

The difference in the slope of the mid-IR SEDs of Arp\,220 and Mrk\,231
is interpreted as evidence for an AGN in the core of 
Mrk\,231. Powerful ionizing radiation near to the AGN would heat a very small 
fraction of the dust grains to high temperatures, boosting the flux density of the 
galaxy at short 
wavelengths and thus producing a shallower SED. There have been various 
attempts to investigate the mid-/far-IR SEDs of 
dust-obscured AGN 
using radiative transfer 
models (Granato et al.\ 1996; Green \& 
Rowan-Robinson 1996), 
including the effects of an obscuring dust torus, the orientation of which 
determines the optical depth to the AGN, and thus its appearance in the  
optical (Antonucci 1993) and mid-infrared wavebands.
 
However, given that a power-law spectrum convincingly accounts for the 
limited mid-IR data available, it is reasonable to avoid such 
details
and assume that 
the SED of the galaxy can be described by the superposition 
of the emission from dust clouds at different temperatures, corresponding to 
different distances from the nucleus in the case that a significant 
fraction of dust heating in a galaxy is due to an AGN. If the dust temperature
is assumed to depend on the distance from the nucleus $r$ as 
$T_{\rm r}(r) \propto r^\eta$ (for a blackbody in equilibrium with an unobscured 
point source $\eta = -0.5$), and the mass of dust enclosed in the spherical shell 
between radii $r$ and $r+{\rm d}r$ is $m_{\rm r}(r) \propto r^\gamma$, then 
dust at each temperature can be associated with a $\delta$-function spectrum, 
$\delta[\nu-\nu_0]$, in which $\nu_0 \simeq (3+\beta) k T / h$ (Blain \& Longair 
1993), and the mid-IR SED, 
\begin{equation} 
f_\nu \propto  \int_{r_{\rm min}}^{r_{\rm max}} m_{\rm r}(r) \, 
T_{\rm r}(r)^{4+\beta} \, \delta[\nu(1+z) - \nu_0] \, {\rm d}r. 
\end{equation}
$\beta$ is the index 
in the function that describes the spectral emissivity of dust 
$\epsilon_\nu \propto \nu^\beta$; $\beta \simeq 1$--1.5. Dust temperatures 
between $T_{\rm min} = 40$\,K and $T_{\rm max} = 2000$\,K correspond to 
emission at wavelengths from about 90 to 1.8\,$\mu$m.
$f_\nu$ thus evaluates to a power law $\nu^p$ with 
$p = 3+\beta + [(\gamma+1)/\eta]$. 
More details of the calculation can be found in Blain (1999).

\begin{table}
\begin{center} 
\caption{The redshift, optical spectral type (Antonucci 1993), 
restframe dust temperature, and mid-IR spectral 
index $p$ -- where $f_\nu \propto \nu^p$ -- for the galaxies whose SEDs 
are plotted in Fig.\,1. For comparison, the spectral index $p \simeq -2.0$ to 
$-2.5$ for samples of low-redshift {\it IRAS} galaxies, either those selected 
at a wavelength of 25\,$\mu$m (Xu et al.\ 1998) or for the most 
luminous selected at 60\,$\mu$m (Sanders \& Mirabel 1996). 
}
\begin{tabular}{ccccc}
\hline
Name & Redshift $z$ & Type & Restframe dust & Mid-IR spectral \\
 & & & temperature $T_{\rm d}$ / K & index $p$ \\
\hline
Arp 220 & 0.02 & 2 & $50 \pm 1$ & $-3.6 \pm 0.2$ \\
Markarian 231 & 0.03 & 1 & $47 \pm 1$ & $-1.9 \pm 0.2$ \\
IRAS\,F10214+4724 & 2.3 & 2 & $76 \pm 4$ & $-1.7 \pm 0.3$ \\
H1413+117 & 2.6 & 1 & $75 \pm 4$ & $-1.1 \pm 0.2$ \\
SMM\,J02399$-$0136 & 2.8 & 2 & $53 \pm 4$ & $-1.7 \pm 0.2$ \\
APM\,08279+5255 & 3.9 & 1 & $107\pm8$ & $-1.0 \pm 0.3$ \\
\hline
\end{tabular}
\end{center} 
\end{table}

Two of the three high-$z$ galaxies with {\it IRAS} detections -- APM\,08279 
and H1413 -- have much flatter mid-IR SEDs as compared with a typical 
low-$z$ {\it IRAS} galaxy or SMM\,J02399: see Fig.\,1. Although SMM\,J02399 
lacks a detection by {\it IRAS}, its mid-IR SED is constrained by a 15-$\mu$m 
{\it ISO} measurement of L. Metcalfe (Ivison et al. 1998). 
Given the limited mid-IR data for SMM\,J02399, it is certainly possible 
that its mid-IR SED could be steeper than the value of $p=-1.7$ listed in Table\,1. 
Another high-$z$ source, the brightest detected in a 850-$\mu$m survey of the 
Hubble Deep Field (Hughes et al.\ 1998) has a reported 15-$\mu$m flux density 
upper limit that indicates a mid-IR SED with $p < -1.7$. Note that the SEDs of 
these galaxies are unlikely to be affected by contaminating sources picked up in 
the different observing beams used at each wavelength. The beam sizes of the 
telescopes are about 5, 10, 25, 40, 7 and 
14\,arcsec at 12, 25, 60, 100, 450 and 850\,$\mu$m respectively. Any 
contamination from other sources nearby on the sky would thus be most 
significant at wavelengths of 
60 and 100\,$\mu$m, and lead to a artificial steepening of the mid-IR SED. 
The low-redshift sources listed in Table\,1 are flagged as unresolved at these 
wavelengths in the {\it IRAS} catalogue, apart from Arp\,220 at 
12\,$\mu$m, and so aperture photometry errors are not expected to
modify their mid-IR SEDs significantly. 

The shallow SEDs of APM\,08279 and H1413 could be due to either an 
unusually large fraction of hot dust in these galaxies, 
or a relatively face-on aspect of 
an obscuring torus, as indicated by a type\,1 optical spectrum, which would 
allow a direct view of the hot dust in the inner regions of the torus. However, 
both are known to be gravitationally lensed, and a systematically 
greater magnification for hotter dust components would also increase the flux 
density at shorter wavelengths and flatten the spectrum (Eisenhardt et al.\ 
1996; Lewis et al.\ 1998).  This situation would arise very naturally if the hotter 
dust clouds are smaller and more central, as described above.

For a large magnification to occur, a distant source must lie close to a 
caustic curve of a gravitational lens. The magnification is formally infinite on 
such a curve, but an upper limit $A_{\rm max}$ is imposed to the magnification 
if the source has a finite size $d$ (Peacock 1982); $A_{\rm max} \propto d^{-1}$. 
This relationship between size and magnification holds regardless of the 
geometry of the source. The diagnostic feature of such a situation is the 
production of multiple images of 
comparable brightness. This is more clearly the case for both APM\,08279 and 
H1413 than for IRAS\,F10214; however, IRAS\,F10214 lies very close to the tip 
of an astroid caustic (Broadhurst \& L\'ehar 1995), and so differential lensing 
would still be expected to flatten its mid-IR SED. 

The effects of differential magnification on the mid-IR SED can be calculated by 
including a factor of $A_{\rm max}(r) \propto r^{-1}$ in the integrand of 
equation (1). In this case, the mid-IR spectral index 
$p = 3+\beta + (\gamma/\eta)$, reduced by $\Delta p = 1/\eta$ as compared with 
the unlensed value. Note that the 
form of the modification to the SED depends only on the value of $\eta$, and 
thus the form of $T(r)$. It is independent of the value of $\gamma$ and the form of 
$m(r)$. Following the same approach, but assuming an exponential decrease of 
$T(r)$, $\Delta p \simeq 1$.

\section{Discussion} 

The mid-IR SEDs of the high- and low-$z$ AGNs SMM\,J02399 and Mrk\,231 
have spectral indices $p \simeq -1.8$. Despite being lensed by a cluster of 
galaxies SMM\,J02399 does not lie close to a caustic, and so is not magnified 
differentially. The strongly lensed distant galaxies APM 08279 and H1413 
have a flatter spectrum with $p \simeq -1.1$. If the difference is caused 
entirely 
by differential magnification,
rather than by a torus orientation effect (Granato et al. 1996),  
then, as $\Delta p (= 1/\eta) \simeq -0.7$, 
$\eta \simeq -1.4$. Reasonably, this value of $\eta$ corresponds to a 
temperature profile that declines more rapidly as compared with that of an  
unscreened blackbody. To match the observed SEDs, $\gamma \simeq 7.3$, 
indicating that most of the emitting dust is relatively cool. 

IRAS\,F10214 has a steeper mid-IR SED, similar to those of Mrk\,231 
and SMM\,J02399, and a much smaller optical flux density as compared with 
APM\,08279 and H1413 (Lewis et al. 1998). These features can both be 
explained if the optical depth of dust extinction into the central regions of 
IRAS\,F10214 is sufficiently large to obscure the AGN from view even in the 
mid-IR waveband, consistent with an inclined optically thick dust torus
and the type-2 optical spectrum of this galaxy. Alternatively, the intrinsic 
unmagnified SED of IRAS\,F10214 could be steeper than those of 
APM\,08279 and H1413, just as these two galaxies could have 
intrinsically flat mid-IR SEDs. 

Does a value of $\eta=-1.4$ correspond to a plausible size for the emitting 
region of a ULIRG? Dust sublimes at a temperature $T \simeq 2000$\,K, 
which is the equilibrium temperature of a blackbody heated by a 
$10^{13}$-L$_\odot$ point source at a distance of 0.6\,parsec. If $\eta = -1.4$, 
then $T$ falls to 100, 50 and 30\,K at distances of 5, 8 and 14\,parsec 
respectively. This is all in agreement with observations of low-$z$ ULIRGs, in 
which the emitting region is less than several hundred parsecs in extent 
(Downes \& Solomon 1998). Most of the luminosity of the galaxy is emitted by 
cool dust on larger scales. The equivalent radius of a 
blackbody sphere emitting 10$^{13}$\,L$_\odot$ at 50\,K is 950\,pc, several 
times larger than the observed sizes of nearby ULIRGs. 

The caustic curves predicted by the lens models that describe the
image configurations observed in IRAS\,F10214 
(Broadhurst \& L\'ehar 1995) and H1413 (Kneib et al.\ 1998) are about 0.8 
and 0.2\,arcsec in size in the plane of the source, larger than the scale of a 
200-pc high-$z$ source. Hence, the whole emitting region of the source 
should be subject to differential magnification; a difference in magnification by a 
factor of about 100 would be obtained between a 200-pc outer radius and a 
2-pc inner radius. 

If the far-IR SED of a galaxy is modeled by a single-temperature dust 
spectrum, then differential magnification would be expected to increase the 
temperature for which the best fit was obtained, by a factor of about 10 to 
20\,per cent 
for discrete data points at the wavelengths given in Fig.\,1. This effect may 
account for at least part of the difference between the very high dust 
temperature of 107\,K inferred for APM\,08279 and the cooler dust temperatures 
generally inferred for other ULIRGs. 
 
The mid-IR SEDs of distant galaxies that are either unlensed or known not 
to suffer from differential magnification are very uncertain. This is likely to 
remain true for some years. In order to probe the spatial structure and 
temperature distribution of dust in the inner regions of AGN directly, a 
space-borne mid-IR interferometer with 500-m baselines at 20\,$\mu$m will 
be required (Mather et al.\ 1998). In the meantime, it might be possible to use 
observations of variability to probe the same regions. The energy output from 
an AGN can vary strongly on short timescales. The response time of the 
hottest dust clouds, on sub-parsec scales, should be comparable to the 
light crossing time -- about 1\,yr. Over the lifetime of 
{\it SIRTF}, any such variations, amplified by differential 
magnification, should be detectable.

\section{Conclusions} 

\begin{enumerate} 
\item 
The slopes of the mid-IR SEDs of high-$z$ lensed ULIRGs are likely to be flatter
than those of unlensed ULIRGs because of differential magnification. While 
observations of these sources must be exploited in order to investigate the 
nature of this class of galaxies, careful account must be taken of the potential 
uncertainties introduced by differential magnification, as well as by the effects 
of the orientation of a dust torus. Recent samples of submillimetre-selected 
galaxies should be immune from this problem. 
\item 
The SED of the longest known high-$z$ ULIRG IRAS\,F10214 is probably 
affected by differential magnification in the same way as those of APM\,08279 
and H1413. However, it has a steeper mid-IR SED. This could be due to either 
a steep intrinsic SED or to a great optical depth, which obscures the most 
central regions even in the mid-IR waveband.  
\item
The SEDs of a sample of lensed galaxies can be used to probe the conditions 
in the cores of distant dusty galaxies. A large sample of these objects will 
be compiled by the {\it Planck Surveyor} satellite (Blain 1998). A direct test 
of the conditions in the central regions of ULIRGs will require a space-borne 
mid-IR interferometer (Mather et al.\ 1998). 
\end{enumerate} 

\section*{Acknowledgements}

I thank the staff of the Observatoire Midi-Pyr\'en\'ees for their 
hospitality, and the MENRT for support while this work was completed.

\end{document}